\documentclass[twocolumn,9pt]{article}
\usepackage{extsizes}
\usepackage[super,sort&compress,comma]{natbib} 
\usepackage[left=1.5cm, right=1.5cm, top=1.785cm, bottom=2.0cm]{geometry}
\usepackage{balance}
\usepackage{mathptmx}
\usepackage{sectsty}
\usepackage{graphicx} 
\usepackage{lastpage}
\usepackage[format=plain,justification=justified,singlelinecheck=false,font={stretch=1.125,small,sf},labelfont=bf,labelsep=space]{caption}
\usepackage{float}
\usepackage{fancyhdr}
\usepackage{fnpos}

\usepackage[english]{babel}
\addto{\captionsenglish}{%
  
}
\usepackage{array}
\usepackage{droidsans}
\usepackage{charter}
\usepackage[T1]{fontenc}
\usepackage[usenames,dvipsnames]{xcolor}
\usepackage{setspace}
\usepackage[compact]{titlesec}
\usepackage{xcolor}
\usepackage{hyperref}
\usepackage{subcaption}
\usepackage{epstopdf}

\newenvironment{changemargin}[2]{%
\begin{list}{}{%
\setlength{\topsep}{0pt}%
\setlength{\leftmargin}{#1}%
\setlength{\rightmargin}{#2}%
\setlength{\listparindent}{\parindent}%
\setlength{\itemindent}{\parindent}%
\setlength{\parsep}{\parskip}%
}
\item[]}{\end{list}}

\title{\LARGE{\textbf{Phase-separation during sedimentation of dilute bacterial suspensions$^\dag$}}}
\date{\today}

\DeclareFontFamily{U}{euc}{}% I chose euc because the chart is called Euler cursive
\DeclareFontShape{U}{euc}{m}{n}{<-6>eurm5<6-8>eurm7<8->eurm10}{}% 
\DeclareSymbolFont{AMSc}{U}{euc}{m}{n} % I chose AMSc because AMSa and AMSb are defined in the amsfonts-package
\DeclareMathSymbol{\umu}{\mathord}{AMSc}{"16}

\begin{document}

\twocolumn[
 \begin{@twocolumnfalse}

\begin{center}
\noindent\LARGE{\textbf{Phase-separation during sedimentation of dilute bacterial suspensions}}\\
\end{center}
\begin{changemargin}{+1.5cm}{+1.5cm}
  \noindent\large{Bryan O. Torres Maldonado$^1$, Ranjiangshang Ran $^1$, K. Lawrence Galloway$^1$, Quentin Brosseau$^1$, Shravan Pradeep$^2$, and Paulo E. Arratia$^{1\dag}$} 
  
 \end{changemargin}
 \begin{changemargin}{+2.2cm}{+2.2cm}
 \textit{$^{1)}$Department of Mechanical Engineering \& Applied Mechanics,
		University of Pennsylvania, Philadelphia, PA 19104.}\\
\textit{%
$^{2)}$Department of Earth and Environmental Science,
		University of Pennsylvania, Philadelphia, PA 19104.\\  $^\dag$E-mail:parratia@seas.upenn.edu%\\This line break forced% with \\
}%\\

 \end{changemargin}
\begin{changemargin}{+1.5cm}{+1.5cm}
\noindent\normalsize{\\Numerous natural systems depend on the sedimentation of passive particles in presence of swimming microorganisms. Here, we investigate the dynamics of the sedimentation of spherical colloids at various \textit{E. coli} concentration within the dilute regime. Results show the appearance of two sedimentation fronts, a spherical particle front and the bacteria front. We find that the bacteria front behave diffusive at short times, whereas at long times decays linearly. The sedimentation speed of passive particles decays at a constant speed and decreases as bacteria concentration ($\phi_b$) is increased. As $\phi_b$ is increased further, the sedimentation speed becomes independent of $\phi_b$. The timescales of the bacteria front is associated with the particle settling speeds. Remarkably, all experiments collapse onto a single master line by using the bacteria front timescale. A phenomenological model is proposed that captures the sedimentation of passive particles in active fluids.} \\
\end{changemargin}
 \end{@twocolumnfalse} ]
 %\footnotetext{\textit{Department of Mechanical Engineering \& Applied Mechanics, University of Pennsylvania, Philadelphia, PA 19104. E-mail: parratia@seas.upenn.edu}}
\section{Introduction}
% Significance:
The  settling  of  particles  in  the  presence  of  microorganisms is a ubiquitous natural process from small river tributaries to lakes to oceans \cite{schallenberg_ecology_1993,zhang_impact_2020,roberto_sediment_2018,nealson_sediment_1997,herndl_microbial_2013, vaccaro_viability_nodate}. Sedimentation of biological matter, for instance, plays an important role on the distribution of plankton in oceans, which is key part of the carbon cycle (i.e. ocean's biological pump) that transports carbon from the ocean's surface to depth \cite{treguer2000global,sarmiento1984model,Ecology_book}. From a technological stand point, microbial activity can affect many process such as the production of food, biofuels, and even vaccines \cite{giorgi_primary_2018,boock_engineered_2019,biom10081184,hugenholtz_nutraceutical_2002,BOUTILIER20094370}. While the settling of passive particles, as a function of suspension volume fraction ($\phi$) and interparticle interactions, have been extensively studied \cite{batchelor1972sedimentation, richardson1954sedimentation, russel1991colloidal, guazzelli2011physical, guazzelli_fluctuations_2011, Piazza_2014, Durian_PhysRevFluids, brady_sedPoF_1988, davis_sedimentation_nodate}, much less is known about particle sedimentation in the presence of swimming microorganisms. 

At infinite dilution, the particle sedimentation speed is set by the balance between the gravitational and viscous drag forces, and the associated velocity called the Stokes' settling speed is given as, $V_0=2a^2\Delta \rho g/9\eta$, where $a$ is the particle radius, $\Delta \rho$ is the density difference between the particle and the solvent, $g$ is the acceleration due to gravity, and $\eta$ is the solvent viscosity \cite{russel1991colloidal}. As $\phi$ increases, the competition between the interparticle forces and thermal motion determines the suspension microstructure, affecting the hydrodynamic interactions between the particles, and in turn the suspension sedimentation speed. These effects are on suspension sedimentation speed are usually described by the hindered settling function, given as $H(\phi)=V_p(\phi)/V_0$, where $V_p(\phi)$ is the mean sedimentation speed due to the upward fluid flow between the settling particles \cite{Durian_PhysRevFluids}.

% Transition to active (steady)
Obtaining the exact form of $H(\phi)$ for quiescent particle suspensions is quite challenging and has been the subject of much investigation \cite{Burger_historical_2001, guazzelli2011physical}. One of the first empirical functional form that captures the $\phi$-dependence on the hindered settling function was proposed by Richardson and Zaki nearly 70 years ago \cite{richardson1954sedimentation}, as $H(\phi)=(1-\phi)^n$ where $n \approx 4.65$ for particles Reynolds numbers ($Re$) below 0.2 when wall effects can be neglected. This relationship was found to be valid for both sedimentation and fluidization processes. It is worth emphasizing that there is still a debate on the exact value of the exponent $n$. Recently, a comprehensive study has shown that the exponent has in fact two values: $n \approx 5.5$ for Brownian particles and $n \approx 4.5$ for non-Brownian particles. At very dilute particle concentration, the exponents converge to $n \approx 6.5$, a value that is close to Batchelor's theoretical prediction, $H(\phi)=1-6.55\phi+\mathcal{O}(\phi^{2})$, obtained by including many-body hydrodynamic interactions between monodisperse constituent particles \cite{batchelor1972sedimentation}. 

% Active Matter Intro
While there is a rich history (and much still to understand) on the subject of sedimentation of passive suspensions \cite{davis_sedimentation_nodate,guazzelli_fluctuations_2011,Piazza_2014,Piazza_2012,Durian_PhysRevFluids,guazzelli2011physical}, there has been a growing interest in systems that emulate the sedimentation of natural processes, particularly those that include living matter such as swimming microorganisms \cite{palacci_sedimentation_2010,hermann_active_2018, D0SM02115F,ginot_nonequilibrium_2015,vachier_dynamics_2019,ginot_sedimentation_2018}. Presence of active particles (living or synthetic) inject energy to a suspending solution. Activity can drive the fluid out of equilibrium even in the absence of external forcing and lead to many poorly understood phenomena including low values of reduced viscosity \cite{lopez_turning_2015,gachelin_non-newtonian_2013}, enhancement in particle diffusivity \cite{wu_particle_2000,chen_fluctuations_2007,mino_enhanced_2011,jepson_enhanced_2013,patteson_particle_2016}, and collective motion (at high densities) \cite{marchetti_review, Berg2008, Ramaswamy_2010_ARCM}. Thus, one expects activity to also affect the sedimentation of passive particles and the associated hindered settling function $H(\phi)$. 

% Intro sed active suspensions (steady)
Experiments with synthetic active Janus particles has shown some intriguing results \cite{palacci_sedimentation_2010,ginot_sedimentation_2018,ginot_nonequilibrium_2015}. These investigations show that the particle \textit{steady-state} density profiles are significantly affected by activity and can be described by an exponential decay \cite{palacci_sedimentation_2010}. Increasing particle activity leads to a slower exponential decay, which is described by using an effective temperature (and diffusivity) that is much larger than for passive systems \cite{palacci_sedimentation_2010}. Varying sedimentation speed while maintaining a constant system activity also led to exponential decay in particle diffusivity \cite{ginot_sedimentation_2018}. These steady-state results agree well with theory and simulations \cite{Cates2010,Cates2008, Tsao2014,vachier_dynamics_2019,ginot_sedimentation_2018}.

% Remaining questions in the field:
Sedimentation experiments with live organisms have been far less common but also show unusual phenomena. For example, bacteria has been shown to enhance sedimentation rates in the presence of polymers due to aggregation \cite{Poon2012}. In mixtures of swimming algae and particles, the (exponential) steady-state sedimentation profile of passive particle is found to be described by an effective diffusivity (or temperature) that increases linearly with the concentration of swimming microbes \cite{Polin2016}, similar to results in artificial active particles. Recently it was found that the sedimentation speed of passive particles can be significantly hindered by the presence of swimming microorganisms (\textit{E. coli}) in the dilute regime \cite{D0SM02115F}; the hindered settling function for these settling active suspensions is captured by a Richardson-Zaki-type equation \cite{richardson1954sedimentation} with an exponent that is much larger than for passive systems and a function of bacteria activity. The \textit{time-dependent} concentration profiles can be adequately described by an advection-diffusion equation coupled with sink-source terms to describe bacteria population dynamics, and with a dispersivity parameter that increases linearly with bacteria concentration \cite{D0SM02115F}. Nevertheless, the effects of bacterial activity on suspension sedimentation dynamics is still not fully understood. 

% Summary of findings presented here: 
Here, we experimentally investigate the sedimentation dynamics of dilute active suspensions using mixtures of passive colloidal spheres and swimming \textit{E. coli}. These initially well-mixed suspensions settle for a period up to three days (depending on bacteria concentration $\phi_{b}$) and image analysis techniques are used to track the height of the particle sedimentation front. We find that the suspension phase separate into a particle- and bacteria-rich fronts. A simple model is presented that can capture the phase separation time based on species dispersivities and settling speeds. 

%
%%%%%%%%END OF INTRODUCTION%%%%%%

\section{Experimental Methods}

% description of suspensions:
In this contribution we experimentally study sedimentation of particles in the presence of bacteria. Active suspensions are prepared by mixing \textit{Escherichia coli} (wild-type K12 MG1655) and polystyrene spheres (density $\rho =1.05$~g/cm${}^3$, Thermo Scientific) of diameter $d = 2$~$\umu$m in deionized (DI) water. Bacteria are grown to saturation ($10^9$ cells/mL) in culture media (LB broth, Sigma-Aldrich). Both spheres and the saturated culture are centrifuged (Centrifuge 5430, eppendorf) and mixed with DI water to reach the desired bacteria volume fraction.
%
%%%%%%%%%%%%%% Figure 1: Schematic and qualitative results%%%%%%%%%%%%%%%%%%%
 \begin{figure}
 \centering
 \includegraphics[height=4.3cm]{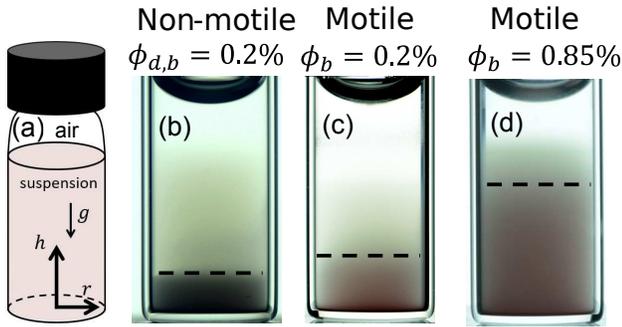}
 \caption{Experimental setup and sample images are shown from left to right: (a) A schematic of the setup and bacteria/particle suspensions. Passive particles are 2~$\umu$m polystyrene spheres, subject to gravity; volume fraction $\phi_p =0.04\%$ kept constant for all experiments. Swimming activity is provided by \textit{E. coli}, a 2~$\umu m$ rod shaped bacterium. Sample snapshots of sedimentation experiments with (b) a mixture of passive particles ($\phi_p =0.04\%$) with non-motile bacteria (\textit{E. coli}) at $\phi_{d,b} =0.20\%$, (c) mixture of passive particles with motile bacteria (\textit{E. coli}) at $\phi_{b} =0.20\%$ and (d) a mixture of passive particles with motile bacteria (\textit{E. coli}) at $\phi_{b} =0.85\%$. All vials samples are shown at time $t=40$~hr. Note the appearance of two sedimentation fronts, bacteria and particle.
 }
 \label{Fig_1}
\end{figure}
%%%%%%%%%%%%%%%%%%%%%%%%%%%%% END OF FIGURE 1%%%%%%%
%
% Intro Experimental Setup

We transfer 1.5~mL of the prepared suspensions into glass vials (9.7~mm in diameter, 35~mm in height), as shown schematically in Fig. \ref{Fig_1}(a); the liquid column is approximately 20~mm tall. The suspensions are homogenized by hand with a pipette so that the bacteria and colloids are uniformly distributed at the start of the experiment. These vials are placed in a water tank to maintain the vials at room temperature $T_{0}=22$~\textdegree C and to remove optical aberrations during imaging (cylindrical vials are preferred to avoid effects from sharp edges). Images are taken every 10~min for a period of 2-3 days with a Nikon D7100 camera that is equipped with a 105~mm Sigma lens. The light source is a Edge-lit Collimated Backlights (CX,Advanced Illumination). 

 % Describe Concentrations
The initial motile bacteria volume fraction $\phi_b$ in the vials ranges from $0.15\%$ to $1.0\%$, while the spherical colloids volume fraction is kept constant at $\phi_p=0.04\%$. Both $\phi_b$ and $\phi_p$ concentrations are considered dilute, and no large scale collective behavior are observed for these bacteria-particle suspensions. The range of $\phi_b$ used here are below which collective motion is generally observed ($\approx 10^{10}$~cells/mL) \cite{kasyap_hydrodynamic_2014}. Under these dilute conditions, \textit{E. coli} swims at speeds ranging from $10$~$\umu$m/s to $20$~$\umu$m/s by the actuation of a helical flagellar bundle. This mode of swimming, under steady conditions, can be described in the far field by an extensile dipole flow (i.e., "pusher") \cite{Berg2008}. 

%%%%%%%%%END OF EXPERIMENTAL METHODS%%
%

\section{Results and Discussion}
%Qualitative Results 
Sample snapshots of the sedimentation experiments are shown in Fig. \ref{Fig_1}(b)$\,$--$\,$(d) at $t=40$~hr. The control passive case, (mixture of non-motile bacteria at $\phi_{d,b}=0.2\%$ with spherical colloidal particles at $\phi_{p}=0.04\%$) is shown in Fig. \ref{Fig_1}(b), while motile bacteria at $\phi_{b}=0.20\%$ and $\phi_{b}=0.85\%$ cases are shown in Fig. \ref{Fig_1}(c) and Fig. \ref{Fig_1}(d), respectively. The images show that bacterial activity hinders the settling speed of passive particles, as previously reported \cite{D0SM02115F}. We find that the passive particle sedimentation front sits higher in the $\phi_{b}=0.20\%$ case than in the non-motile bacteria case ($\phi_{d,b}=0.20\%$). This trend continues for the $\phi_{b}=0.85\%$ case, which indicates that particle settling speed decreases with increasing $\phi_{b}$ (see SM for more information). We also observe the formation of two sedimentation fronts, one that is rich in passive particle and one that is rich in bacteria, as shown in Fig. \ref{Fig_1}(b)$\,$--$\,$(d). These two fronts settle at different speeds. These qualitative results show that (i) bacteria significantly hinders particle sedimentation speed and (ii) the two species phase separate during the sedimentation process. While the appearance of the bacteria front has been previously reported \cite{D0SM02115F}, here we will focus on quantifying and understanding the appearance of the two sedimentation fronts and its relationship to particle sedimentation front speed. 
%%%%%%%%%%%%%%%%% Figure 2: Kymograph%%%%%%%%%%%%%%%%
 \begin{figure}
 \centering
 \includegraphics[height=6.6cm]{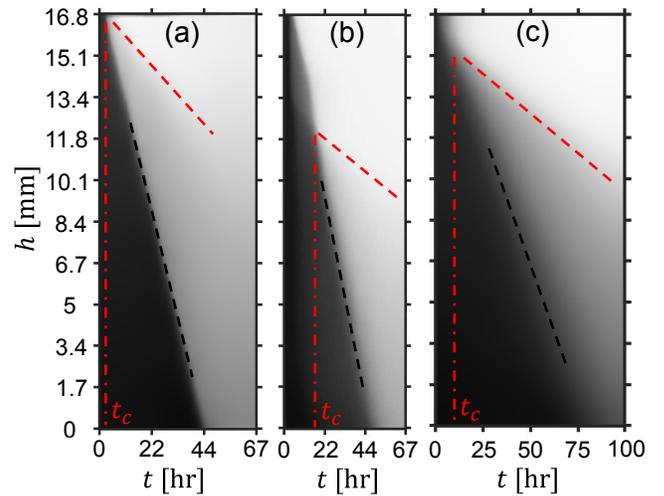}
 \caption{
Kymographs for sedimentation at fixed particle concentration ($\phi_p =0.04\%$) and different bacterial concentrations. (a) With non-motile bacteria $\phi_{d,b} =0.20\%$, (b) motile bacteria at $\phi_{b} =0.20\%$, and (c) motile bacteria at $\phi_{b} =0.85\%$. Two fronts are seen when $\phi_b \geq 0.15\%$. Note that activity hinders the settling speed of spherical particles.
  }
 \label{Fig_2}
\end{figure}
%%%%%%%%%%%%%%%%%%%%%%%%%%%%%%%%%%% END OF FIGURE 2%%%%%%%%%%%%%%%%%%%%%%%

%Introducing second front with V_{l,b} and $V_{d,b}=0.04 um/s

\subsection{Front Sedimentation Heights \& Speeds}
We begin by measuring the sedimentation height of the passive spheres front as a function of time at different $\phi_b$. The particle front sedimentation height, $h_p(t)$, is located in the vial by identifying the peak in the vertical gradient of the image intensity. By plotting $h_p(t)$ as a function of time, we can extract the sedimentation speed of the particle front $V_p$ and define a hindered settling function $H(\phi_b)$ as a function of bacteria concentration. The hindered settling function is defined as the ratio of the mean sedimentation speed $V_p$ of the particle front over the sedimentation speed of a single particle such that $H(\phi_b)=V_p(\phi_b)/V_0$. %The sedimentation speed of a single polystyrene colloid $V_0$ in a viscous fluid of viscosity $\eta$ in absence of inertia is calculated by the force balance between the gravitational force and viscous drag acting on the particle. This yields  $V_0= \Delta \rho g d^2/{18 \eta}$, where $\Delta \rho$ is the density difference between the colloid and the suspending liquid, $g$ is the acceleration due to gravity ($g = 9.81$ m/s${}^2$), $\eta$ is the fluid viscosity, and $d$ is the particle diameter. 
We note that the Stokes' sedimentation speed $V_0$ is $0.12~\umu$m/s for the 2 $\umu$m polystyrene particles in water.\\
%Kymographs and introducing t_C for motile and non-motile bacteria
 %%%%%%%%%%%%%Inserting figure 3%%%%%%%%%%%%%%%%%%%%%%
\begin{figure}[h!t]
 \centering
 \includegraphics[height=6.2cm]{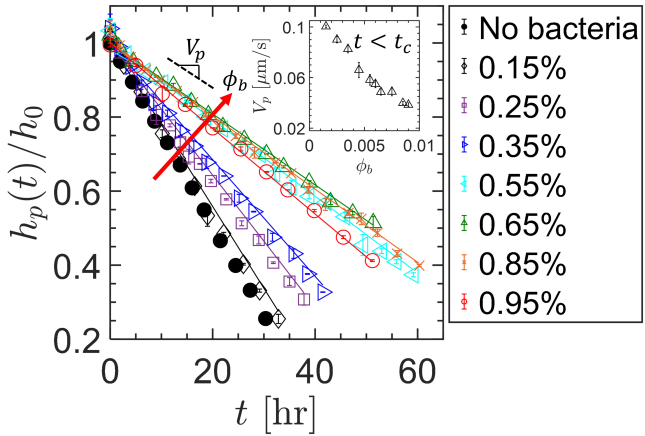}
 \caption{Height of particle sedimentation front, $h_p(t)/h_0$, where $h_0$ is the initial height tracked on the sedimentation process as a function of time for a range of initial motile bacterial volume fractions $\phi_b$. Data shows the decrease in front speed with an increase of bacteria volume fraction $\phi_b$. %(a) At long times, 
 The slope of $h_p(t)/h_0$ does not vary significantly when $\phi_b>0.40\%$. Solid lines are obtained by solving Equation~\ref{ode_eq}. Solution to this expression is shown in Equation~\ref{ode_sol}, and it captures relatively well the height of the front of passive particles for a wide range of bacterial concentrations $\phi_b$. Inset shows  particle front speed ($V_p$) as a function of $\phi_b$ for short times ($t<t_c$).
 }
 \label{Fig_3}
\end{figure}
%%%%%%%%%%%%%End figure 3%%%%%%%%%%%%%%%%%%
% Introducing Kymograph
Figure \ref{Fig_2}(a)$\,$--$\,$(c) shows sedimentation kymographs ($h$ vs $t$) for the mixture of particles with non-motile bacteria ($\phi_{d,b}=0.20\%$) case and two with active suspensions, $\phi_{b}=0.20\%$ and $\phi_{b}=0.85\%$, respectively; these kymographs correspond to the data shown in Fig. \ref{Fig_1}(b)$\,$--$\,$(d). These experimental kymographs are obtained by the juxtaposition of a single pixel line at the center of the vial for each image of the time series. As noted before, all samples have a fixed particle volume fraction ($\phi_p =0.04\%$). In each case a sedimentation front is formed and moves downward from the top of the container at a speed $V_p$. A first look at the corresponding kymographs reveal that the velocity $V_p$, here visible as the slope of the colloid front (black dashed line), decreases as the concentration of bacteria increases. We again observe the formation of a bacteria front that separates from spherical particles after a time $t_c$ (red dashed line). The timescale $t_c$ is defined as the time that the bacteria-rich front diverge from the particle-rich front. Experiments with non-motile bacteria and spherical particles shows that the two phases separate at relatively short time-scales ($t<1$~hr), as shown in Fig. \ref{Fig_2}(a). Suspensions of motile bacteria (with particles) on the other hand phase separate at much longer time-scales ($t>10$~hr). That is, activity seems to delay phase separation. Moreover, the phase separation time, $t_c$, seems to decrease with $\phi_b$, that is, $t_c=t_c(\phi_b)$. 

% Comparing results of passive case (phi_b=0)with single sphere 
The particle front sedimentation speed $V_p$ is found by linear fits to the kymograph plots such that 
 \begin{equation}
 h_p(t)= h_0-V_p t,
   	\label{hp_eq}
\end{equation}
where $h_0$ is the initial front position; in all cases studied here, this linear relationship works relatively well even at long times ($t>24$~hr). In absence of bacteria ($\phi_b=0\%$), the front sedimentation speed $V_p=0.11~\umu$m/s, which is relatively close to the sedimentation speed of a single sphere, $V_0=0.12~\umu$m/s. This result indicates that hydrodynamic interaction among passive particles are relatively weak in our study. This not surprising since particle volume fraction is quite dilute, $\phi_p= 0.04\%$.

% settling is linear, and changes with bacteria
Figure \ref{Fig_3} shows the normalized height of the particle sedimentation front, $h(t)/h_0$, for different bacteria volume fractions $\phi_b$, where $h_0$ is the front initial height. We find that the settling height of the particle front is linear through all experiments, so it is described by a average speed, $V_p$. A constant sedimentation speed is not a trivial result because the decrease in bacteria activity is commensurable with the time of the experiment. At 48~hr of experiment, approximately 70$\%$ of the population has lost motility and behave as a passive colloid \cite{D0SM02115F}. This has been previously characterized by the bacteria motility loss rate $k$, which is approximately $6\times 10^{-6}~\mathrm{s}^{-1}$ (see Ref. \cite{D0SM02115F}). We believe that the decrease in activity promotes a faster sedimentation; concurrently, the increase in passive particles (non-motile bacteria) should decrease the overall sedimentation speed. These two processes are not necessarily of the same magnitude.

% Here We talk about the two regimes by looking the fronts
The data also show that $V_p$ decreases as $\phi_b$ increases (Fig. \ref{Fig_3}). Surprisingly, this trend weakens and $V_p$ becomes independent of $\phi_b$ for $\phi_b>0.40\%$ at long times ($t>t_c$). This transition, however, is not observed at short times ($t<t_c$). To better illustrate this trend, we plot $V_p$ as a function of $\phi_b$ for all cases (Fig. \ref{Fig_3}, inset); we find that, for $t<t_c$, $V_p$ decreases linearly with $\phi_b$. Overall, we find two different functionalities regarding the effects of bacteria activity on $V_p$: (i) for $t<t_c$, $V_p$ decreases linearly with $\phi_b$ for all cases; (ii) for $t>t_c$, $V_p$ decreases linearly with $\phi_b$ for $\phi_b \leq 0.4\%$ followed by an asymptote for $\phi_b \geq 0.45\%$.  

\subsection{Hindered Settling Function for Active Suspensions}

We now compute the suspension hindered settling function $H$ as a function of bacteria concentration, $\phi_b$; note that $\phi_p$ is kept constant at 0.04\% for all cases. This variable is explored as a touchstone to previous literature and as a tool to develop our phenomenological model for the sedimentation of active suspensions. Similar to the $V_p$ data (and as expected), we find two regimes for $H(\phi_b)$. (i) For $t<t_c$, $H$ decreases linearly with $\phi_b$ for all cases [Fig. \ref{Fig_4}(a), black triangles]; (ii) for $t>t_c$, $H$ decreases linearly with $\phi_b$ for $\phi_b \leq 0.4\%$ followed by an asymptote ($H \approx 0.45$) for $\phi_b \geq 0.45\%$ [Fig. \ref{Fig_4}(a), red diamonds]. These data show that bacterial activity can reduce the particle sedimentation speed by a factor of 2, which illustrates the dramatic effect of bacterial activity on sedimentation dynamics. The first regime ($t<t_c$) can be described by a linearized version of the Richardson-zaki equation \cite{richardson1954sedimentation} such that $H(\phi_b)=(1-n\phi_b)$, with a best fit of $n=95$ [Fig. \ref{Fig_4}(a), black line]. This exponent is significantly larger than the expected fitting parameter for Brownian particles ($n \approx 5.5$) \cite{Durian_PhysRevFluids} and Batchelor's result ($n=6.5$) \cite{batchelor1972sedimentation}, but in line with our own previous results ($n \approx 120$) \cite{D0SM02115F}. While the mechanisms leading to the linear dependence of $H$ on $\phi_b$ was previously investigated in our previous work \cite{D0SM02115F}, the transition of $H$ from a linear dependence to a $\phi_b$-independent regime at $t>t_c$ has yet to be reported and/or understood. 

The observed $\phi_b$-independent regime in $H$ may be due to the increase in the amount (or volume fraction) of non-motile bacteria within the suspension during the sedimentation process. If we assume a a constant motility loss rate $k$ \cite{D0SM02115F}, it follows that an increase in $\phi_b$ would result in the production of more non-motile bacteria for the same experimental time $\Delta t$. This increase in non-motile concentration within the system may overwhelm the effect of activity ($\phi_b$), although it is not enough to significantly affect particle sedimentation speed. That is, the dependence of $H$ on passive (non-motile) particle volume fraction is much weaker than on motile bacterial volume fraction. 

\subsection{Phase Separation During Sedimentation}% A possible mechanism for H(phi_b)

%%%%%%%%%%%%%%%%%%%Inserting figure 4%%%%%%%%%%%%%%%%%%%%%
\begin{figure}
 \centering
 \includegraphics[height=10cm]{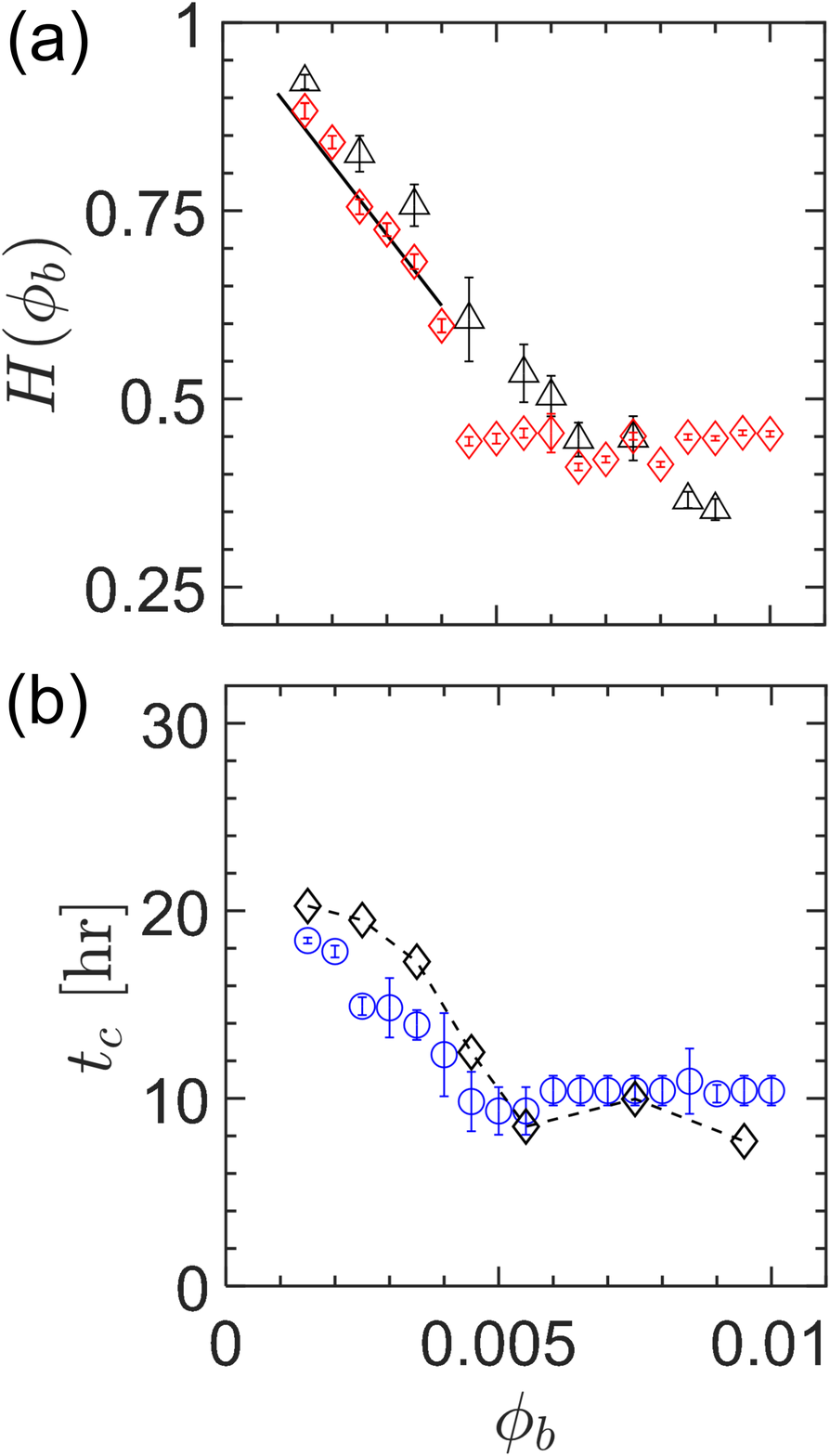}
 \caption{(a) Hindered settling function $H(\phi_b)$ as a function of bacterial volume fraction $\phi_b$. $H(\phi_b)$ was obtained at short times (black triangles) and at long times (red diamonds). At short times ($t<t_c$) we observed that sedimentation is hindered linearly in the experiments. However, at long times ($t>t_c$), $H$ decreases linearly with $\phi_b$ for $\phi_b \leq 0.4$ followed by an asymptote ($H \approx 0.45$) for $\phi_b > 0.45$ (b) Phase separation times-scale $t_c$ (blue circles) as a function of bacteria volume fraction $\phi_b$. Solid line is the times-scale prediction obtained by Equation~\ref{tc_eq} with a time dependent diffusivity that decays with time. Experimental results shows that the timescale $t_c$ has the same functionality as $H(\phi_b)$. %Time of oxygen depletion $t{O_2}$ (red square) is quantified as a function of $\phi_b$, which overestimates $t_c$, but it has similar trends to experimental measurements of $t_c$.
 }   
 \label{Fig_4}
\end{figure}
%%%%%%%%%%%%%%%%END OF FIGURE 4%%%%%%%%%%%%%%%
% Time-scale t_c: microscopic
The images and kymographs shown in Fig. \ref{Fig_1} and Fig. \ref{Fig_2}, respectively, show that after a certain period of sedimentation time, the active suspension separates into particle-rich and bacteria-rich fronts. Figure \ref{Fig_4}(b) [blue circles] shows the experimentally measured phase separation time $t_c$ as a function of $\phi_b$ for all cases. Similar to $H(\phi_b)$, the quantity $t_c$ shows a decreasing trend with $\phi_b$, suggesting a deeper underlying relation at play. We now provide a simple argument to understand the origins of $t_c$ by noting that the behavior of the bacteria front is diffusive and can be captured by an equation of the form:
\begin{equation}
 h_b(t)= h_0-V_b t-2 \sqrt{D_b t}.
   	\label{hb_eq}
\end{equation}
Here, $D_b$ is a dispersivity related to bacterial activity and $V_b$ is the sedimentation speed of the bacteria front. A dispersion length scale can thus be defined as $L_D(t)=2 \sqrt{D_b t}$. At earlier times, the bacteria front [or Eq.~\ref{hb_eq}] settles faster than the particle front [or Eq.~\ref{hp_eq}], but slower at later times (see SM Fig. 3). The phase separation time, $t_c$, is the time at which the particle front crosses over the bacteria front, or $h_b(t_c)= h_p(t_c)$. By equating Eq.~\ref{hp_eq} and Eq.~\ref{hb_eq}, we obtain the following equation for the phase separation time: 
\begin{equation}
	t_c=\frac{4D_b}{(V_p - V_b)^2}.
   	\label{tc_eq}
\end{equation}
All variables in the above equation can be measured or estimated. For example, $V_p$ is experimentally measured as a function of $\phi_b$ (Fig. 3) and $V_{b}=0.02\pm 0.002~\umu$m/s is approximately independent of $\phi_b$ in the dilute regime investigated here (see SM for more information). The dispersivity $D_b$ can be obtained by fitting Eq.~\ref{hb_eq} to the bacteria front shown in the kymograph. The data shows that $D_b$ decays exponentially with time as such that $D_b(t)=D_0 \exp(-\beta t)$, where $D_0$ is the initial dispersivity, and $\beta$ is a dispersivity decay rate; the values of $D_0$ and $\beta$ can be found in the SM. The decay in dispersivity over time may be due to nutrient and oxygen depletion, or even aerotactic response near the air-liquid interface \cite{PhysRevLett.102.198101,SCHWARZLINEK20162} (see SM for more information). For simplicity, we will proceed with an average dispersivity $\overline{D_b}$ defined as:
\begin{equation}
    \overline{D_b}=\frac{1}{t_0}\int_{0}^{t_0} D_b(t)\,dt, 
\end{equation}

We plot the calculated phase separation time using $\overline{D_b}$ in Fig. ~\ref{Fig_4}(b) [black diamonds]. Results show that this simple argument is able to capture the nonlinear experimental data relatively well, including the decreasing trend of $t_c$ with $\phi_b$. Here, $t_0$ is a long enough time to provide a statistically meaningful average, longer than the experimentally measured phase separation time.

\subsubsection{Scaling for Hindering Settling Function}
\begin{figure}[h!t]
 \centering
 \includegraphics[height=10cm]{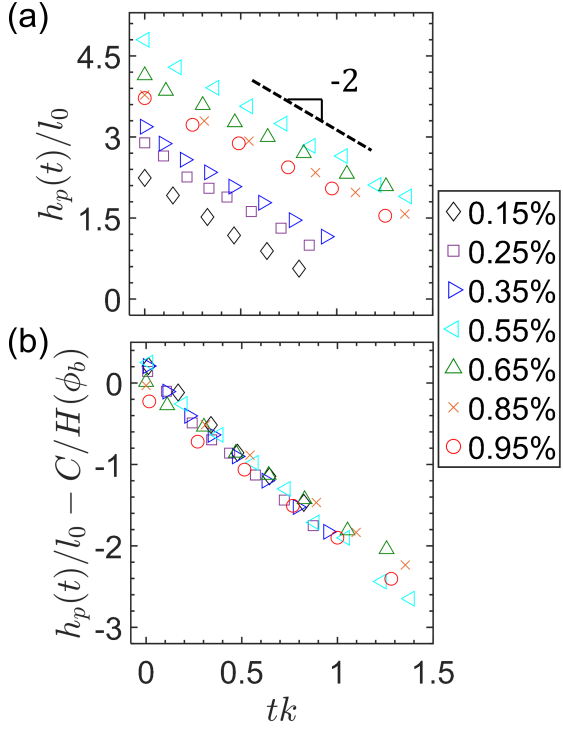}
 \caption{(a) Normalized height $h_p(t)/l_0$ as a function of normalized time $t k$. Length-scale $l_0$ is defined as $l_0=V_0 t_c$. Normalized plot shows a constant slope on all experiments, the slope as a value of $1.97 \pm 0.050$. (b) Normalized height as a function of the normalized time. The height was initially normalized by the length-scale $l_0$, and the time by the motility loss rate $k$. This normalization gave a constant slope of $1.97 \pm 0.050$ as shown in (a). Now, the normalized height is shifted by a constant $C$ divided by the hindered settling function $H(\phi_b)$. The shift on the normalized height of the experimental data gives a collapse into a single line at different $\phi_b$. The constant has a value of $C=1.955\pm 0.045$.
 }
 \label{Fig_5}
\end{figure}

The data show above makes it clear that there is a relationship between the timescale associated with phase separations $t_c$ and the hindering settling function $H(\phi_b)$. The other time scale present in the experiments is the bacteria motility loss rate, $k$. We can use these two time-scales, $t_c$ and $k$, to non-dimensionalize both $h_p(t)$ and $t$. The quantity $t_c$ is used to obtain a "separation" length-scale, $l_0 = t_c V_0$, to normalize $h_p(t)$ such that $h_p(t)/l_0$. This quantity tells us how far or close the front is from separating. Next, we non-dimensionalize time $t$ with the motility loss rate, $k$. The quantity $tk$ provides a measure of activity left in the sample. Figure \ref{Fig_5}(a) shows the normalized particle sedimentation height, $h_p(t)/l_0$, as a function of scaled time, $tk$, for a range of $\phi_b$. The data shows that $h_p(t)/l_0$ decreases linearly with $tk$, suggesting that activity prevents phase separation likely by mixing the two phases during the sedimentation process. Also, we find an average slope of $-1.97 \pm 0.050$ through all experiments, indicating that these sample share similar dynamics. 

%%%%%%%%%%%%%Inserting figure 5 %%%%%%%%%%%%%%
%%%%%%%%%%%%End figure 5%%%%%%%%%%%%%%%%%
% capture the intercept: 
Next, we account for the intercept in Fig. \ref{Fig_5}(a) by noticing that it changes similarly to the inverse of the hindered settling function, $H(\phi_b$) (see SM for more information). Therefore, we subtract the re-scaled height ($h_p(t)/l_0$) by a constant divided by the hindered settling function such that $C/H(\phi_b)$. The constant $C$ is obtained by equating the intercept of $h_p(t)/l_0$ with $C/H(\phi_b)$, giving a value of $1.955 \pm 0.045$. Remarkably, all experiments collapse onto a single master curve [Fig. \ref{Fig_5}(b)]. Putting together these results, the sedimentation front of passive particles can be described by the following differential equation:
\begin{equation}
	\frac{h_p(t)}{t_c V_0} - C\frac{V_0}{dh/dt} = -2kt,
	\label{ode_eq}
\end{equation}
which can be rearranged to be
\begin{equation}
	\frac{dh_p(t)}{dt}\left(\frac{h_p(t)}{t_c V_0}+2kt\right)=C V_0. 
	\label{ode_eq_alt}
\end{equation}
We take the linear solution to this equation to obtain
\begin{equation}
	h_p(t)=-2k t_c V_0 t +\frac{C V_0}{2k}.
	\label{ode_sol}
\end{equation}
% summary of terms: 
Equation \ref{ode_eq} shows that the sedimentation front of the passive particles are described by three main terms. The first term is (sedimentation) height normalized by a diffusion-based length-scale associated with the bacteria-particle settling process. The second term quantifies passive settling under gravity via the hindered settling function. The third term is time normalized by the motility decay rate, and quantifies the level of activity left in the sample. We note that a solution to Equation \ref{ode_eq} is a line and is shown in Equation \ref{ode_sol}. If we solve the equation and revisit front height over time [Fig. \ref{Fig_3}], we observe that the solution captures all cases of active suspensions.

Our experimental observations show that sedimentation and separation of passive particles and bacteria suspensions are described by using sedimentation and diffusive length scales. The quantity $t_c$ of active suspensions occurs over very long time-scales compared to mixtures of passive particles that have two different masses and is reminiscent of the effects of centrifuging. This timescale plateau above a critical bacteria concentration ($ \phi_b = 0.40 \%$). We find that when $t_c$ plateau when $\phi_b>0.40 \%$, and $H(\phi_b)$ has the same functionality. A possible mechanism that explains our experimental observations is due to an increased presence of non-motile bacteria which overwhelm the effects of the increased activity. The injection of energy by the bacteria swimming motion plays an important role in the sedimentation process; its decay is quantified by a motility loss rate $k$, which is used to describe the settling of spherical particles. 

%wrap up: discus future work
Preparation of new experiments that track spherical particles while they sediment would help to understand these observations. This could be achieved by using a confocal microscope with a smaller vial. The downside of this proposed approach is that controlling the temperature at long times is a challenge. With this approach we would be able to obtain 3-D imaging as a function of time, which would allow to track the spherical particles and explore particle fluctuations.

%%%%%%%%% END OF RESULTS AND DISCUSSION %%%%%%%%%%
%%%%%%%%%%%%%%% CONCLUSIONS %%%%%%%%%%%%%%%

\section{Conclusions}
The sedimentation of passive particles in the presence of swimming bacteria is experimentally investigated. We find that bacterial activity (even in the dilute regime) hinders the sedimentation of passive particles and phase separates at large time-scales, whereas the bacteria front with non-motile bacteria shows at smaller time-scales ($t<1$~hr). Particle hindered settling function, $H(\phi_b$), as well as $t_c$, show two distinct regimes: (i) a monotonic decay by increasing $\phi_b$ and (ii) a regime in which both $H(\phi_b$) and $t_c$ are no longer dependent of $\phi_b$. A simple model based on particle and bacteria length scales is proposed which describes $t_c$ for a range of $\phi_b$. Using non-dimensional analysis, we collapse all sedimentation front data onto a single master line that captures both bacteria concentration regimes. A phenomenological model is proposed to describe the height of the particle sedimentation front. The solution of this model captures relatively well the height of the sedimentation front of the passive particles for all $\phi_b$ within the dilute regime. In summary, an addition of a diffusive length-scale can captures the short time behavior of the bacteria front, and a sedimentation speed length scales captures the long time behavior of the bacteria front and the particle front. Further study on this field with different colloid concentrations and swimmer actuation modes (pusher vs puller) would presumably shed light on how the hydrodynamic interactions between the passive particles and active bacteria underlay the dynamics of our system and be generalized to others.

%%% END OF Conclusions %%

\section*{Conflicts of interest}
There are no conflicts to declare.

\section*{Acknowledgements}
We thank D. Jerolmack, C. Kammer, and A. Patteson for fruitful discussions. S.P. and P.E.A. acknowledge support from Army Research Office (ARO, Grant W911NF2010113). B.O.T.M., R.R., and P.E.A. acknowledge support by the National Science Foundation grant DMR-1709763.\\
\bibliographystyle{rsc}
\bibliography{sedimentation}

\providecommand*{\mcitethebibliography}{\thebibliography}
\csname @ifundefined\endcsname{endmcitethebibliography}
{\let\endmcitethebibliography\endthebibliography}{}
\begin{mcitethebibliography}{49}
\providecommand*{\natexlab}[1]{#1}
\providecommand*{\mciteSetBstSublistMode}[1]{}
\providecommand*{\mciteSetBstMaxWidthForm}[2]{}
\providecommand*{\mciteBstWouldAddEndPuncttrue}
  {\def\EndOfBibitem{\unskip.}}
\providecommand*{\mciteBstWouldAddEndPunctfalse}
  {\let\EndOfBibitem\relax}
\providecommand*{\mciteSetBstMidEndSepPunct}[3]{}
\providecommand*{\mciteSetBstSublistLabelBeginEnd}[3]{}
\providecommand*{\EndOfBibitem}{}
\mciteSetBstSublistMode{f}
\mciteSetBstMaxWidthForm{subitem}
{(\emph{\alph{mcitesubitemcount}})}
\mciteSetBstSublistLabelBeginEnd{\mcitemaxwidthsubitemform\space}
{\relax}{\relax}

\bibitem[Schallenberg and Kalff(1993)]{schallenberg_ecology_1993}
M.~Schallenberg and J.~Kalff, \emph{Ecology}, 1993, \textbf{74}, 919--934\relax
\mciteBstWouldAddEndPuncttrue
\mciteSetBstMidEndSepPunct{\mcitedefaultmidpunct}
{\mcitedefaultendpunct}{\mcitedefaultseppunct}\relax
\EndOfBibitem
\bibitem[Zhang \emph{et~al.}(2020)Zhang, Tu, Li, Lu, and Li]{zhang_impact_2020}
L.~Zhang, D.~Tu, X.~Li, W.~Lu and J.~Li, \emph{BMC Microbiology}, 2020,
  \textbf{20}, 254\relax
\mciteBstWouldAddEndPuncttrue
\mciteSetBstMidEndSepPunct{\mcitedefaultmidpunct}
{\mcitedefaultendpunct}{\mcitedefaultseppunct}\relax
\EndOfBibitem
\bibitem[Roberto \emph{et~al.}(2018)Roberto, Van~Gray, and
  Leff]{roberto_sediment_2018}
A.~A. Roberto, J.~B. Van~Gray and L.~G. Leff, \emph{Water Research}, 2018,
  \textbf{134}, 353--369\relax
\mciteBstWouldAddEndPuncttrue
\mciteSetBstMidEndSepPunct{\mcitedefaultmidpunct}
{\mcitedefaultendpunct}{\mcitedefaultseppunct}\relax
\EndOfBibitem
\bibitem[Nealson(1997)]{nealson_sediment_1997}
K.~H. Nealson, \emph{Annual Review of Earth and Planetary Sciences}, 1997,
  \textbf{25}, 403--434\relax
\mciteBstWouldAddEndPuncttrue
\mciteSetBstMidEndSepPunct{\mcitedefaultmidpunct}
{\mcitedefaultendpunct}{\mcitedefaultseppunct}\relax
\EndOfBibitem
\bibitem[Herndl and Reinthaler(2013)]{herndl_microbial_2013}
G.~J. Herndl and T.~Reinthaler, \emph{Nature Geoscience}, 2013, \textbf{6},
  718--724\relax
\mciteBstWouldAddEndPuncttrue
\mciteSetBstMidEndSepPunct{\mcitedefaultmidpunct}
{\mcitedefaultendpunct}{\mcitedefaultseppunct}\relax
\EndOfBibitem
\bibitem[Vaccaro \emph{et~al.}(1950)Vaccaro, Briggs, Carey, and
  Ketchum]{vaccaro_viability_nodate}
R.~F. Vaccaro, M.~P. Briggs, L.~Carey and B.~H. Ketchum, \emph{American Journal
  of Public Health}, 1950, \textbf{40}, 1257--1266\relax
\mciteBstWouldAddEndPuncttrue
\mciteSetBstMidEndSepPunct{\mcitedefaultmidpunct}
{\mcitedefaultendpunct}{\mcitedefaultseppunct}\relax
\EndOfBibitem
\bibitem[Tr{\'e}guer and Pondaven(2000)]{treguer2000global}
P.~Tr{\'e}guer and P.~Pondaven, \emph{Nature}, 2000, \textbf{406}, 358\relax
\mciteBstWouldAddEndPuncttrue
\mciteSetBstMidEndSepPunct{\mcitedefaultmidpunct}
{\mcitedefaultendpunct}{\mcitedefaultseppunct}\relax
\EndOfBibitem
\bibitem[Sarmiento and Toggweiler(1984)]{sarmiento1984model}
J.~Sarmiento and J.~Toggweiler, \emph{Nature}, 1984, \textbf{308},
  621--624\relax
\mciteBstWouldAddEndPuncttrue
\mciteSetBstMidEndSepPunct{\mcitedefaultmidpunct}
{\mcitedefaultendpunct}{\mcitedefaultseppunct}\relax
\EndOfBibitem
\bibitem[Kiørboe(2008)]{Ecology_book}
T.~Kiørboe, \emph{A Mechanistic Approach to Plankton Ecology}, Princeton
  University Press, 2008\relax
\mciteBstWouldAddEndPuncttrue
\mciteSetBstMidEndSepPunct{\mcitedefaultmidpunct}
{\mcitedefaultendpunct}{\mcitedefaultseppunct}\relax
\EndOfBibitem
\bibitem[Giorgi \emph{et~al.}(2018)Giorgi, Reitsma, van Fulpen, Berg, and
  Bechger]{giorgi_primary_2018}
S.~Giorgi, B.~A.~H. Reitsma, H.~J.~F. van Fulpen, R.~W.~P. Berg and M.~Bechger,
  \emph{Water Science and Technology}, 2018, \textbf{78}, 1597--1602\relax
\mciteBstWouldAddEndPuncttrue
\mciteSetBstMidEndSepPunct{\mcitedefaultmidpunct}
{\mcitedefaultendpunct}{\mcitedefaultseppunct}\relax
\EndOfBibitem
\bibitem[Boock \emph{et~al.}(2019)Boock, Freedman, Tompsett, Muse, Allen,
  Jackson, Castro-Dominguez, Timko, Prather, and
  Thompson]{boock_engineered_2019}
J.~T. Boock, A.~J.~E. Freedman, G.~A. Tompsett, S.~K. Muse, A.~J. Allen, L.~A.
  Jackson, B.~Castro-Dominguez, M.~T. Timko, K.~L.~J. Prather and J.~R.
  Thompson, \emph{Nature Communications}, 2019, \textbf{10}, 587\relax
\mciteBstWouldAddEndPuncttrue
\mciteSetBstMidEndSepPunct{\mcitedefaultmidpunct}
{\mcitedefaultendpunct}{\mcitedefaultseppunct}\relax
\EndOfBibitem
\bibitem[Vázquez \emph{et~al.}(2020)Vázquez, Durán, Menduíña, and
  Nogueira]{biom10081184}
J.~A. Vázquez, A.~I. Durán, A.~Menduíña and M.~Nogueira,
  \emph{Biomolecules}, 2020, \textbf{10}, \relax
\mciteBstWouldAddEndPuncttrue
\mciteSetBstMidEndSepPunct{\mcitedefaultmidpunct}
{\mcitedefaultendpunct}{\mcitedefaultseppunct}\relax
\EndOfBibitem
\bibitem[Hugenholtz and Smid(2002)]{hugenholtz_nutraceutical_2002}
J.~Hugenholtz and E.~J. Smid, \emph{Current Opinion in Biotechnology}, 2002,
  \textbf{13}, 497--507\relax
\mciteBstWouldAddEndPuncttrue
\mciteSetBstMidEndSepPunct{\mcitedefaultmidpunct}
{\mcitedefaultendpunct}{\mcitedefaultseppunct}\relax
\EndOfBibitem
\bibitem[Boutilier \emph{et~al.}(2009)Boutilier, Jamieson, Gordon, Lake, and
  Hart]{BOUTILIER20094370}
L.~Boutilier, R.~Jamieson, R.~Gordon, C.~Lake and W.~Hart, \emph{Water
  Research}, 2009, \textbf{43}, 4370--4380\relax
\mciteBstWouldAddEndPuncttrue
\mciteSetBstMidEndSepPunct{\mcitedefaultmidpunct}
{\mcitedefaultendpunct}{\mcitedefaultseppunct}\relax
\EndOfBibitem
\bibitem[Batchelor(1972)]{batchelor1972sedimentation}
G.~Batchelor, \emph{Journal of fluid mechanics}, 1972, \textbf{52},
  245--268\relax
\mciteBstWouldAddEndPuncttrue
\mciteSetBstMidEndSepPunct{\mcitedefaultmidpunct}
{\mcitedefaultendpunct}{\mcitedefaultseppunct}\relax
\EndOfBibitem
\bibitem[Richardson(1954)]{richardson1954sedimentation}
J.~Richardson, \emph{Transactions of the institution of chemical engineers},
  1954, \textbf{32}, 35--53\relax
\mciteBstWouldAddEndPuncttrue
\mciteSetBstMidEndSepPunct{\mcitedefaultmidpunct}
{\mcitedefaultendpunct}{\mcitedefaultseppunct}\relax
\EndOfBibitem
\bibitem[Russel \emph{et~al.}(1991)Russel, Russel, Saville, and
  Schowalter]{russel1991colloidal}
W.~B. Russel, W.~Russel, D.~A. Saville and W.~R. Schowalter, \emph{Colloidal
  dispersions}, Cambridge university press, 1991\relax
\mciteBstWouldAddEndPuncttrue
\mciteSetBstMidEndSepPunct{\mcitedefaultmidpunct}
{\mcitedefaultendpunct}{\mcitedefaultseppunct}\relax
\EndOfBibitem
\bibitem[Guazzelli and Morris(2011)]{guazzelli2011physical}
E.~Guazzelli and J.~F. Morris, \emph{A physical introduction to suspension
  dynamics}, Cambridge University Press, 2011, vol.~45\relax
\mciteBstWouldAddEndPuncttrue
\mciteSetBstMidEndSepPunct{\mcitedefaultmidpunct}
{\mcitedefaultendpunct}{\mcitedefaultseppunct}\relax
\EndOfBibitem
\bibitem[Guazzelli and Hinch(2011)]{guazzelli_fluctuations_2011}
E.~Guazzelli and J.~Hinch, \emph{Annual Review of Fluid Mechanics}, 2011,
  \textbf{43}, 97--116\relax
\mciteBstWouldAddEndPuncttrue
\mciteSetBstMidEndSepPunct{\mcitedefaultmidpunct}
{\mcitedefaultendpunct}{\mcitedefaultseppunct}\relax
\EndOfBibitem
\bibitem[Piazza(2014)]{Piazza_2014}
R.~Piazza, \emph{Reports on Progress in Physics}, 2014, \textbf{77},
  056602\relax
\mciteBstWouldAddEndPuncttrue
\mciteSetBstMidEndSepPunct{\mcitedefaultmidpunct}
{\mcitedefaultendpunct}{\mcitedefaultseppunct}\relax
\EndOfBibitem
\bibitem[Brzinski and Durian(2018)]{Durian_PhysRevFluids}
T.~A. Brzinski and D.~J. Durian, \emph{Phys. Rev. Fluids}, 2018, \textbf{3},
  124303\relax
\mciteBstWouldAddEndPuncttrue
\mciteSetBstMidEndSepPunct{\mcitedefaultmidpunct}
{\mcitedefaultendpunct}{\mcitedefaultseppunct}\relax
\EndOfBibitem
\bibitem[Brady(1988)]{brady_sedPoF_1988}
J.~F. Brady, \emph{The Physics of fluids}, 1988, \textbf{31}, 717--727\relax
\mciteBstWouldAddEndPuncttrue
\mciteSetBstMidEndSepPunct{\mcitedefaultmidpunct}
{\mcitedefaultendpunct}{\mcitedefaultseppunct}\relax
\EndOfBibitem
\bibitem[Davis and Acrivos(1985)]{davis_sedimentation_nodate}
R.~H. Davis and A.~Acrivos, \emph{Annual Review of Fluid Mechanics}, 1985,
  \textbf{17}, 91--118\relax
\mciteBstWouldAddEndPuncttrue
\mciteSetBstMidEndSepPunct{\mcitedefaultmidpunct}
{\mcitedefaultendpunct}{\mcitedefaultseppunct}\relax
\EndOfBibitem
\bibitem[Bürger and Wendland(2001)]{Burger_historical_2001}
R.~Bürger and W.~L. Wendland, \emph{Journal of Engineering Mathematics}, 2001,
  \textbf{41}, 101--116\relax
\mciteBstWouldAddEndPuncttrue
\mciteSetBstMidEndSepPunct{\mcitedefaultmidpunct}
{\mcitedefaultendpunct}{\mcitedefaultseppunct}\relax
\EndOfBibitem
\bibitem[Piazza \emph{et~al.}(2012)Piazza, Buzzaccaro, and Secchi]{Piazza_2012}
R.~Piazza, S.~Buzzaccaro and E.~Secchi, \emph{Journal of Physics: Condensed
  Matter}, 2012, \textbf{24}, 284109\relax
\mciteBstWouldAddEndPuncttrue
\mciteSetBstMidEndSepPunct{\mcitedefaultmidpunct}
{\mcitedefaultendpunct}{\mcitedefaultseppunct}\relax
\EndOfBibitem
\bibitem[Palacci \emph{et~al.}(2010)Palacci, Cottin-Bizonne, Ybert, and
  Bocquet]{palacci_sedimentation_2010}
J.~Palacci, C.~Cottin-Bizonne, C.~Ybert and L.~Bocquet, \emph{Physical Review
  Letters}, 2010, \textbf{105}, 088304\relax
\mciteBstWouldAddEndPuncttrue
\mciteSetBstMidEndSepPunct{\mcitedefaultmidpunct}
{\mcitedefaultendpunct}{\mcitedefaultseppunct}\relax
\EndOfBibitem
\bibitem[Hermann and Schmidt(2018)]{hermann_active_2018}
S.~Hermann and M.~Schmidt, \emph{Soft Matter}, 2018, \textbf{14},
  1614--1621\relax
\mciteBstWouldAddEndPuncttrue
\mciteSetBstMidEndSepPunct{\mcitedefaultmidpunct}
{\mcitedefaultendpunct}{\mcitedefaultseppunct}\relax
\EndOfBibitem
\bibitem[Singh \emph{et~al.}(2021)Singh, Patteson, Torres~Maldonado, Purohit,
  and Arratia]{D0SM02115F}
J.~Singh, A.~Patteson, B.~Torres~Maldonado, P.~K. Purohit and P.~E. Arratia,
  \emph{Soft Matter}, 2021,  4151--4160\relax
\mciteBstWouldAddEndPuncttrue
\mciteSetBstMidEndSepPunct{\mcitedefaultmidpunct}
{\mcitedefaultendpunct}{\mcitedefaultseppunct}\relax
\EndOfBibitem
\bibitem[Ginot \emph{et~al.}(2015)Ginot, Theurkauff, Levis, Ybert, Bocquet,
  Berthier, and Cottin-Bizonne]{ginot_nonequilibrium_2015}
F.~Ginot, I.~Theurkauff, D.~Levis, C.~Ybert, L.~Bocquet, L.~Berthier and
  C.~Cottin-Bizonne, \emph{Physical Review X}, 2015, \textbf{5}, 011004\relax
\mciteBstWouldAddEndPuncttrue
\mciteSetBstMidEndSepPunct{\mcitedefaultmidpunct}
{\mcitedefaultendpunct}{\mcitedefaultseppunct}\relax
\EndOfBibitem
\bibitem[Vachier and Mazza(2019)]{vachier_dynamics_2019}
J.~Vachier and M.~G. Mazza, \emph{The European Physical Journal E}, 2019,
  \textbf{42}, 11\relax
\mciteBstWouldAddEndPuncttrue
\mciteSetBstMidEndSepPunct{\mcitedefaultmidpunct}
{\mcitedefaultendpunct}{\mcitedefaultseppunct}\relax
\EndOfBibitem
\bibitem[Ginot \emph{et~al.}(2018)Ginot, Solon, Kafri, Ybert, Tailleur, and
  Cottin-Bizonne]{ginot_sedimentation_2018}
F.~Ginot, A.~Solon, Y.~Kafri, C.~Ybert, J.~Tailleur and C.~Cottin-Bizonne,
  \emph{New Journal of Physics}, 2018, \textbf{20}, 115001\relax
\mciteBstWouldAddEndPuncttrue
\mciteSetBstMidEndSepPunct{\mcitedefaultmidpunct}
{\mcitedefaultendpunct}{\mcitedefaultseppunct}\relax
\EndOfBibitem
\bibitem[López \emph{et~al.}(2015)López, Gachelin, Douarche, Auradou, and
  Clément]{lopez_turning_2015}
H.~M. López, J.~Gachelin, C.~Douarche, H.~Auradou and E.~Clément,
  \emph{Physical Review Letters}, 2015, \textbf{115}, 028301\relax
\mciteBstWouldAddEndPuncttrue
\mciteSetBstMidEndSepPunct{\mcitedefaultmidpunct}
{\mcitedefaultendpunct}{\mcitedefaultseppunct}\relax
\EndOfBibitem
\bibitem[Gachelin \emph{et~al.}(2013)Gachelin, Miño, Berthet, Lindner,
  Rousselet, and Clément]{gachelin_non-newtonian_2013}
J.~Gachelin, G.~Miño, H.~Berthet, A.~Lindner, A.~Rousselet and E.~Clément,
  \emph{Physical Review Letters}, 2013, \textbf{110}, 268103\relax
\mciteBstWouldAddEndPuncttrue
\mciteSetBstMidEndSepPunct{\mcitedefaultmidpunct}
{\mcitedefaultendpunct}{\mcitedefaultseppunct}\relax
\EndOfBibitem
\bibitem[Wu and Libchaber(2000)]{wu_particle_2000}
X.-L. Wu and A.~Libchaber, \emph{Physical Review Letters}, 2000, \textbf{84},
  3017--3020\relax
\mciteBstWouldAddEndPuncttrue
\mciteSetBstMidEndSepPunct{\mcitedefaultmidpunct}
{\mcitedefaultendpunct}{\mcitedefaultseppunct}\relax
\EndOfBibitem
\bibitem[Chen \emph{et~al.}(2007)Chen, Lau, Hough, Islam, Goulian, Lubensky,
  and Yodh]{chen_fluctuations_2007}
D.~T.~N. Chen, A.~W.~C. Lau, L.~A. Hough, M.~F. Islam, M.~Goulian, T.~C.
  Lubensky and A.~G. Yodh, \emph{Physical Review Letters}, 2007, \textbf{99},
  148302\relax
\mciteBstWouldAddEndPuncttrue
\mciteSetBstMidEndSepPunct{\mcitedefaultmidpunct}
{\mcitedefaultendpunct}{\mcitedefaultseppunct}\relax
\EndOfBibitem
\bibitem[Miño \emph{et~al.}(2011)Miño, Mallouk, Darnige, Hoyos, Dauchet,
  Dunstan, Soto, Wang, Rousselet, and Clement]{mino_enhanced_2011}
G.~Miño, T.~E. Mallouk, T.~Darnige, M.~Hoyos, J.~Dauchet, J.~Dunstan, R.~Soto,
  Y.~Wang, A.~Rousselet and E.~Clement, \emph{Physical Review Letters}, 2011,
  \textbf{106}, 048102\relax
\mciteBstWouldAddEndPuncttrue
\mciteSetBstMidEndSepPunct{\mcitedefaultmidpunct}
{\mcitedefaultendpunct}{\mcitedefaultseppunct}\relax
\EndOfBibitem
\bibitem[Jepson \emph{et~al.}(2013)Jepson, Martinez, Schwarz-Linek, Morozov,
  and Poon]{jepson_enhanced_2013}
A.~Jepson, V.~A. Martinez, J.~Schwarz-Linek, A.~Morozov and W.~C.~K. Poon,
  \emph{Physical Review E}, 2013, \textbf{88}, 041002\relax
\mciteBstWouldAddEndPuncttrue
\mciteSetBstMidEndSepPunct{\mcitedefaultmidpunct}
{\mcitedefaultendpunct}{\mcitedefaultseppunct}\relax
\EndOfBibitem
\bibitem[Patteson \emph{et~al.}(2016)Patteson, Gopinath, Purohit, and
  Arratia]{patteson_particle_2016}
A.~E. Patteson, A.~Gopinath, P.~K. Purohit and P.~E. Arratia, \emph{Soft
  Matter}, 2016, \textbf{12}, 2365--2372\relax
\mciteBstWouldAddEndPuncttrue
\mciteSetBstMidEndSepPunct{\mcitedefaultmidpunct}
{\mcitedefaultendpunct}{\mcitedefaultseppunct}\relax
\EndOfBibitem
\bibitem[Marchetti \emph{et~al.}(2013)Marchetti, Joanny, Ramaswamy, Liverpool,
  Prost, Rao, and Simha]{marchetti_review}
M.~C. Marchetti, J.~F. Joanny, S.~Ramaswamy, T.~B. Liverpool, J.~Prost, M.~Rao
  and R.~A. Simha, \emph{Rev. Mod. Phys.}, 2013, \textbf{85}, 1143--1189\relax
\mciteBstWouldAddEndPuncttrue
\mciteSetBstMidEndSepPunct{\mcitedefaultmidpunct}
{\mcitedefaultendpunct}{\mcitedefaultseppunct}\relax
\EndOfBibitem
\bibitem[Berg(2008)]{Berg2008}
H.~C. Berg, \emph{E. coli in Motion}, Springer Science \& Business Media,
  2008\relax
\mciteBstWouldAddEndPuncttrue
\mciteSetBstMidEndSepPunct{\mcitedefaultmidpunct}
{\mcitedefaultendpunct}{\mcitedefaultseppunct}\relax
\EndOfBibitem
\bibitem[Ramaswamy(2010)]{Ramaswamy_2010_ARCM}
S.~Ramaswamy, \emph{Annual Review of Condensed Matter Physics}, 2010,
  \textbf{1}, 323--345\relax
\mciteBstWouldAddEndPuncttrue
\mciteSetBstMidEndSepPunct{\mcitedefaultmidpunct}
{\mcitedefaultendpunct}{\mcitedefaultseppunct}\relax
\EndOfBibitem
\bibitem[Nash \emph{et~al.}(2010)Nash, Adhikari, Tailleur, and
  Cates]{Cates2010}
R.~W. Nash, R.~Adhikari, J.~Tailleur and M.~E. Cates, \emph{Phys. Rev. Lett.},
  2010, \textbf{104}, 258101\relax
\mciteBstWouldAddEndPuncttrue
\mciteSetBstMidEndSepPunct{\mcitedefaultmidpunct}
{\mcitedefaultendpunct}{\mcitedefaultseppunct}\relax
\EndOfBibitem
\bibitem[Tailleur and Cates(2008)]{Cates2008}
J.~Tailleur and M.~Cates, \emph{Phys. Rev. Lett.}, 2008, \textbf{100},
  218103\relax
\mciteBstWouldAddEndPuncttrue
\mciteSetBstMidEndSepPunct{\mcitedefaultmidpunct}
{\mcitedefaultendpunct}{\mcitedefaultseppunct}\relax
\EndOfBibitem
\bibitem[Wang \emph{et~al.}(2014)Wang, Chen, Sheng, and Tsao]{Tsao2014}
Z.~Wang, H.~Y. Chen, Y.~J. Sheng and H.~K. Tsao, \emph{Soft Matter}, 2014,
  \textbf{10}, 3209--3217\relax
\mciteBstWouldAddEndPuncttrue
\mciteSetBstMidEndSepPunct{\mcitedefaultmidpunct}
{\mcitedefaultendpunct}{\mcitedefaultseppunct}\relax
\EndOfBibitem
\bibitem[Schwarz-Linek \emph{et~al.}(2012)Schwarz-Linek, Valeriani, Cacciuto,
  Cates, Marenduzzo, Morozov, and Poon]{Poon2012}
J.~Schwarz-Linek, C.~Valeriani, A.~Cacciuto, M.~E. Cates, D.~Marenduzzo, A.~N.
  Morozov and W.~C.~K. Poon, \emph{Proc. Natl. Acad. Scien.}, 2012,
  \textbf{109}, 4052--4057\relax
\mciteBstWouldAddEndPuncttrue
\mciteSetBstMidEndSepPunct{\mcitedefaultmidpunct}
{\mcitedefaultendpunct}{\mcitedefaultseppunct}\relax
\EndOfBibitem
\bibitem[Jeanneret \emph{et~al.}(2016)Jeanneret, Pushkin, Kantsler, and
  Polin]{Polin2016}
R.~Jeanneret, D.~O. Pushkin, V.~Kantsler and M.~Polin, \emph{Nature Comm.},
  2016, \textbf{7}, 12518\relax
\mciteBstWouldAddEndPuncttrue
\mciteSetBstMidEndSepPunct{\mcitedefaultmidpunct}
{\mcitedefaultendpunct}{\mcitedefaultseppunct}\relax
\EndOfBibitem
\bibitem[Kasyap \emph{et~al.}(2014)Kasyap, Koch, and
  Wu]{kasyap_hydrodynamic_2014}
T.~V. Kasyap, D.~L. Koch and M.~Wu, \emph{Physics of Fluids}, 2014,
  \textbf{26}, 081901\relax
\mciteBstWouldAddEndPuncttrue
\mciteSetBstMidEndSepPunct{\mcitedefaultmidpunct}
{\mcitedefaultendpunct}{\mcitedefaultseppunct}\relax
\EndOfBibitem
\bibitem[Douarche \emph{et~al.}(2009)Douarche, Buguin, Salman, and
  Libchaber]{PhysRevLett.102.198101}
C.~Douarche, A.~Buguin, H.~Salman and A.~Libchaber, \emph{Phys. Rev. Lett.},
  2009, \textbf{102}, 198101\relax
\mciteBstWouldAddEndPuncttrue
\mciteSetBstMidEndSepPunct{\mcitedefaultmidpunct}
{\mcitedefaultendpunct}{\mcitedefaultseppunct}\relax
\EndOfBibitem
\bibitem[Schwarz-Linek \emph{et~al.}(2016)Schwarz-Linek, Arlt, Jepson, Dawson,
  Vissers, Miroli, Pilizota, Martinez, and Poon]{SCHWARZLINEK20162}
J.~Schwarz-Linek, J.~Arlt, A.~Jepson, A.~Dawson, T.~Vissers, D.~Miroli,
  T.~Pilizota, V.~A. Martinez and W.~C. Poon, \emph{Colloids and Surfaces B:
  Biointerfaces}, 2016, \textbf{137}, 2--16\relax
\mciteBstWouldAddEndPuncttrue
\mciteSetBstMidEndSepPunct{\mcitedefaultmidpunct}
{\mcitedefaultendpunct}{\mcitedefaultseppunct}\relax
\EndOfBibitem
\end{mcitethebibliography}

\end{document}